\documentclass[iop]{emulateapj}

\usepackage{amssymb,amsmath}
\usepackage{mathrsfs}
\usepackage{natbib}
\usepackage{mathtools}
\usepackage{graphicx}
\usepackage{multirow}
\usepackage{subfigure}
\usepackage{eucal}
\usepackage{url}
\usepackage{color}
\usepackage{float}

\usepackage{color}
\definecolor{red}{rgb}{1,0,0}

\begin{document}

\title{Does the Milky Way Obey Spiral Galaxy Scaling Relations?}
\author{Timothy C. Licquia\altaffilmark{1,2}, Jeffrey A. Newman\altaffilmark{1,2}, Matthew A. Bershady\altaffilmark{3}}
\altaffiltext{1}{Department of Physics and Astronomy, University of Pittsburgh, 3941 O'Hara Street, Pittsburgh, PA 15260; {tcl15@pitt.edu, janewman@pitt.edu} }
\altaffiltext{2}{Pittsburgh Particle physics, Astrophysics, and Cosmology Center (PITT PACC)}
\altaffiltext{3}{Department of Astronomy, University of Wisconsin, 475 N. Charter St., Madison, WI 53706, USA; mab@astro.wisc.edu}

\begin{abstract}
It is crucial to understand how the Milky Way, the galaxy we can study in the most intimate detail, fits in amongst other galaxies.  Key examples include the Tully-Fisher relation (TFR) --- i.e., the tight correlation between luminosity ($L$) and rotational velocity ($V_\textrm{rot}$) --- and the 3-dimensional luminosity-velocity-radius ($LVR$) scaling relation.  Several past studies have characterized the MW as a 1--1.5$\sigma$ outlier to the TFR.  This study reexamines such comparisons using new estimates of MW properties that are robust to many of the systematic uncertainties that have been a problem in the past and are based on assumptions consistent with those used for other spiral galaxies.  Comparing to scaling relations derived from modern extragalactic data, we find that our Galaxy's properties are in excellent agreement with TFRs defined using any SDSS-filter absolute magnitude, stellar mass, or baryonic mass as the $L$ proxy.  We next utilize disk scale length ($R_\textrm{d}$) measurements to extend this investigation to the $LVR$ relation.  Here we find that our Galaxy lies farther from the relation than $\sim$90\% of other spiral galaxies, yielding $\sim$9.5$\sigma$ evidence that it is unusually compact for its $L$ and $V_\textrm{rot}$ (based on MW errors alone), a result that holds for all of the $L$ proxies considered.  The expected $R_\textrm{d}$ for the MW from the $LVR$ relation is $\sim$5 kpc, nearly twice as large as the observed value, with error estimates placing the two in tension at the $\sim$1.4$\sigma$ level.  The compact scale length of the Galactic disk could be related to other ways in which the MW has been found to be anomalous.
\end{abstract}


\section{Introduction} \label{sec:intro}
It is well established that spiral galaxies follow a set of power-law relationships amongst their global properties, generally referred to as scaling relations \citep[][hereafter H12]{CourteauRix99, Courteau07, Dutton07, Dutton11, Hall12}. The most prominent of these, known as the Tully-Fisher relation \citep[TFR;][]{TullyFisher77}, reflects the tight relationship between intrinsic brightness (or luminosity, $L$) and rotational velocity ($V_{\rm rot}$). In the standard picture, spiral galaxies generically contain a luminous disk component comprised of rotationally supported stars and gas residing at the center of a much larger and more massive dark matter ÒhaloÓ component \citep{Fall80,Kauffmann93,Cole94,Mo98}. The TFR encodes key information about the ratio of luminous to dark matter in these systems, making it a powerful tool for studying galaxy formation and evolution \citep[e.g.,][]{Dutton07, Dutton11}, though our theoretical understanding of the origins of this relation within a $\Lambda$CDM universe is still progressing (\citealp{Dutton07, Dutton11}; H12). Correlations also are observed between disk size, or radius ($R$), and $L$ or $V_{\rm rot}$, but display significantly larger dispersions (\citealp{Courteau07}; H12).

Determining how the properties of our Galaxy, the Milky Way (MW), fit with these scaling relations is a challenging task. A handful of studies have performed such an analysis, indicating that the MW is a 1--1.5$\sigma$ outlier from the TFR \citep{Malhotra96,Flynn,Hammer07}. This begs the question: is our Galaxy truly atypical, or have we failed to make an apples-to-apples comparison?  The latter case is hard to rule out, as it is impossible for astronomers to measure the MW in the same manner as extragalactic objects, due in large part to our uniquely inside-out perspective. Adding further complications, dust in the interstellar medium obscures most of the MW's stars from view, especially at optical wavelengths \citep{Green15}. Any attempt to infer global properties of our Galaxy directly typically requires extrapolating from measurements of only the most nearby stars --- a tiny fraction of the total stellar population --- using MW models that rest upon a large array of assumptions \citep[e.g.,][]{Malhotra96, Flynn, vdK86, Siegel02, Robin03, Girardi05, Juric08, Lopez14, Mao15}. As a result, estimates of MW properties have remained poorly constrained and vulnerable to systematic errors, making it difficult to assess accurately how our Galaxy fits amongst its peers. 

This study aims at overcoming these challenges, which is important for a number of reasons.  For instance, if the MW is truly representative of spiral galaxies, then our high-detail studies of it may be fairly used to understand the TFR, as well as other driving mechanisms of galaxy evolution.  On the other hand, if our Galaxy does not fit the common mold, then identifying its unusual characteristics could offer important clues into how its formation history may have been unusual (see, e.g., \citealp{Hammer07}; \citealp{Liu11}; \citealp{Busha11MCs}).

This paper is structured as follows.  In \S\ref{sec:data_SDSS} we discuss the extragalactic dataset that we use to measure spiral galaxy scaling relations. In \S\ref{sec:data_mw} we next discuss the MW measurements employed in this study and the methods used to produce them. We then focus on comparing the MW to TFRs in \S\ref{sec:tfr}, with our methodologies outlined in \S\ref{sec:methods_tfr} and our results discussed in \S\ref{sec:results_tfr}.  We next extend this type of comparison to 3-dimensional luminosity-velocity-radius ($LVR$) relations in \S\ref{sec:lvr}, where likewise our methodologies are first outlined in \S\ref{sec:methods_lvr} and our results are then discussed in \S\ref{sec:results_lvr}.  In \S\ref{sec:discussion} we provide broader discussions of our results; this includes the important differences between this study and prior ones that deemed the MW a 1--1.5$\sigma$ outlier to the TFR in \S\ref{sec:discussion_prior_tfr}.  In \S\ref{sec:discussion_toosmall} we discuss how our results lend circumstantial evidence towards the emergent picture of a ``too-small'' MW Galaxy; this includes potential concerns about this conclusion in \S\ref{sec:discussion_toosmall_concerns}, but also how it is supported by studies of the MW's satellite galaxy population in \S\ref{sec:discussion_toosmall_support}. Lastly, we conclude with a summary of our findings in \S\ref{sec:summary}.

Throughout this paper, we use $L$ when generically discussing any proxy for luminosity; these include both absolute magnitudes, denoted by (italicized) $M$, and mass properties, denoted by (unitalicized) M.  In order to compare measurements for extragalactic objects (which rely on assumptions about cosmic distance scales) directly to those for the MW (within which distances may be measured more directly), we adopt a standard $\Lambda$CDM cosmology parameterized by a mass density of dark and normal matter of $\Omega_{\rm M} = 0.3$, an effective mass density of dark energy of $\Omega_\Lambda = 0.7$, and a Hubble constant of $H_0 = 70$ km s$^{-1}$ Mpc$^{-1}$. All absolute magnitudes, on the other hand, are derived using a Hubble constant of $H_0 = 100h$ km s$^{-1}$ Mpc$^{-1}$, therefore making them truly measurements of $M - 5\log h$.  Following convention, all SDSS $ugriz$ magnitudes are reported in the AB system, whereas all Johnson-Cousins $UBVRI$ magnitudes are reported in the Vega system.  In all cases, we use ``log'' to denote the base-10 logarithm.

\section{Data} \label{sec:data} 

In our earlier studies we have developed improved estimates of MW properties that are robust to many sources of systematic uncertainty. Most notably, this includes new determinations of the MW's photometric properties (i.e., its global color and luminosity at optical wavelengths), not by extrapolating from optical measurements in the Solar neighborhood, but by analyzing the integrated properties of \emph{Milky Way analogs} --- a sample of galaxies whose distributions of total stellar masses and star formation rates match the inferred posterior distributions for those same properties of the MW --- observed by the Sloan Digital Sky Survey (SDSS; \citealp{Licquia2}, hereafter L15). We also have utilized a Bayesian mixture-model analysis technique to improve constraints on a variety of other Galactic properties, including the total stellar mass and star formation rate \citep[][hereafter LN15]{Licquia1}, as well as the disk scale length \citep[][in press, hereafter LN16]{Licquia3}. In this paper, we use these results to explore how the MW fits with spiral galaxy scaling relations. In Table \ref{table:mw}, we list the estimates for each of the MW properties that we employ in this study; these reflect the best-to-date values available in the literature and account for all major known sources of uncertainty. All MW values in Table \ref{table:mw} have been adjusted to use the same basic assumptions as the extragalactic scaling relation data used here. Consequentially, the Galactic-to-extragalactic comparisons presented below should be robust to many potential sources of systematic offsets.  We discuss these data more in the following subsections.

\begin{deluxetable*}{lccl}
\tablenum{1} \label{table:mw}
\tablewidth{\textwidth}
\tablecaption{Global Properties of the Milky Way}
\tablehead{Property & Mean$\pm$1$\sigma$ & Units & Description}
\startdata
$V_\text{rot}$ & $220\pm22$ & km s$^{-1}$ & Rotational velocity of the disk \\ 
$R_{\rm d}$ & $2.71^{+0.22}_{-0.20}$ & kpc & Photometric disk scale length \\
$\text{M}_\star$ & $5.67^{+1.53}_{-1.11}\times10^{10}$ & $\text{M}_\odot$ & Total stellar mass \\
$\text{M}_\text{HI}$ & $5.55\pm1.74\times10^{9}$ & $\text{M}_\odot$ & Neutral hydrogen gas mass \\
$\text{M}_\text{bar}$ & $6.45^{+1.54}_{-1.14}\times10^{10}$ & $\text{M}_\odot$ & Baryonic mass ($=\text{M}_\star+1.4\text{M}_\text{HI}$) \\
$^0\!M_u$ & $-19.15^{+0.55}_{-0.47}$ & mag & SDSS $u$-band absolute magnitude \\
$^0\!M_g$ & $-20.33^{+0.42}_{-0.43}$ & mag & SDSS $g$-band absolute magnitude \\ 
$^0\!M_r$ & $-20.97^{+0.37}_{-0.40}$ & mag & SDSS $r$-band absolute magnitude \\
$^0\!M_i$ & $-21.24^{+0.37}_{-0.38}$ & mag & SDSS $i$-band absolute magnitude \\
$^0\!M_z$ & $-21.53^{+0.36}_{-0.39}$ & mag & SDSS $z$-band absolute magnitude \\
$^0\!M_U$ & $-20.00^{+0.59}_{-0.47}$ & mag & Johnson $U$-band absolute magnitude \\
$^0\!M_B$ & $-20.05^{+0.41}_{-0.45}$ & mag & Johnson $B$-band absolute magnitude \\
$^0\!M_V$ & $-20.71^{+0.39}_{-0.40}$ & mag & Johnson $V$-band absolute magnitude \\
$^0\!M_R$ & $-21.23^{+0.39}_{-0.39}$ & mag & Cousins $R$-band absolute magnitude \\
$^0\!M_I$  & $-21.81^{+0.38}_{-0.38}$ & mag & Cousins $I$-band absolute magnitude
\enddata
\tablecomments{See \S\ref{sec:data_mw} for the details on the source of each estimate.  All absolute magnitudes are derived using a cosmic distance scale with Hubble constant $H_0 = 100 h$ km s$^{-1}$ Mpc$^{-1}$, and hence truly represent values of $^0\!M - 5\log h$; the 0 superscript prefix before absolute magnitudes indicates that they include K-corrections to the $z=0$ rest-frame passbands.  The rotational velocity is adopted from \citet{IAU} and we have ascribed a 10\% measurement error.  The estimates listed for $\text{M}_\text{HI}$ and $\text{M}_\text{bar}$ are derived in \S\ref{sec:data_mw}.  The remaining entries come from LN16.}
\end{deluxetable*}

\subsection{SDSS Spiral Galaxies} \label{sec:data_SDSS} 

Our sample of spiral galaxies consists of objects that have both photometric and spectroscopic measurements from the Eighth Data Release \citep[DR8;][]{DR8} of the SDSS \citep{York2000}, as well as 21-cm radio spectral line measurements from the SFI++ catalog \citep{Springob07}. We have identified a sample of 422 spiral galaxies suitable for this study by cross-matching objects from the data used in L15 to those found in both the SFI++ catalog and Table 1 from H12, yielding the following measurements:
\begin{itemize}
\item Spectroscopic redshifts ($z$) with errors typically $<0.1\%$ from SDSS-DR8;
\item SDSS-DR8 $ugriz$ absolute magnitudes that include $K$-corrections to the $z=0$ rest frame ($^0\!M - 5\log h$), based on extinction-corrected {\tt cmodel} photometry, produced in L15;
\item Total stellar masses (M$_\star$) produced in L15 by applying MPA/JHU catalog algorithms \citep{Brinchmann04} to DR8 photometry assuming a Kroupa initial mass function \citep[IMF;][]{Kroupa03};
\item Neutral hydrogen (HI) gas masses (M$_{\rm HI}$) determined from 21-cm line flux measurements that are corrected for beam attenuation, pointing offsets, and HI self-absorption from SFI++ \citep{Springob07};
\item Disk rotational velocities ($V_{\rm rot}$) determined from 21-cm line width measurements that are corrected for instrumental effects, redshift broadening, disk inclination, and distance from SFI++; and
\item Disk inclinations corrected for projection effects and angular scale lengths ($\Theta_{\rm d}$) measured in units of arcseconds by H12 from an isophotal analysis of each galaxy's SDSS $i$-band light profile.
\end{itemize}

We note that H12 performed their own reduction and photometry of galaxy images and demonstrated that their methods yield much more reliable estimates of $\Theta_{\rm d}$ than the SDSS imaging pipeline. Using the measurements listed above we also calculate the baryonic mass of each galaxy as ${\rm M}_{\rm bar} = {\rm M}_\star + 1.4{\rm M}_{\rm HI}$, where the factor of 1.4 accounts for the mass present in the form of helium and metals (following H12), and its physical disk scale length as $R_{\rm d} = \Theta_{\rm d} d_{\rm A}(z)$, where $d_{\rm A}(z)$ is the angular diameter distance at redshift $z$ from our adopted cosmology.

Identified in this way, our sample of 422 objects is analogous to ``Sample C'' of H12, and constitutes the subset of those objects that are found to have SDSS-DR8 imaging and spectroscopy, allowing us to use values from L15 for properties in the first three bullet points above. We next restrict to only those objects with inclinations measured to be in the range of 40 to 75 degrees. This cut eliminates galaxies whose disks are too face-on to have an accurately measured $V_{\rm rot}$ or are too edge-on to ensure an accurately determined $\Theta_{\rm d}$. As a result, we are left with 258 spiral galaxies, analogous to ``Sample D'' of H12, that should be best for determining scaling relations and which we discuss exclusively hereafter. In practice, we find that this inclination restriction reduces the observed scatter in our TFRs by $\sim$10\%; however, the conclusions we draw are entirely insensitive to comparing the MW to scaling relations based on the sample with or without this cut applied (see Tables 2 and \ref{table:LVR}).

\subsection{The Milky Way} \label{sec:data_mw} 

In this section we discuss each of the MW estimates listed in Table \ref{table:mw}. We note that the measurements from our previous work were derived specifically to enable comparative studies such as this one.  That is, our MW estimates rest upon the same basic assumptions that were used for extragalactic objects. Most importantly, these include a Kroupa IMF, a single-exponential disk model, and a consistent definition of stellar mass, which includes the contributions from both main-sequence stars and remnants but not substellar material, following the same convention as the stellar evolution models of \citet{BC03}.

In LN15 we have presented a Bayesian mixture-model (BMM) analysis framework for statistically combining various estimates of MW properties.  In that study, we addressed the question: given the measurements available in the literature, what is the best aggregate estimate for both the Galactic bulge+bar mass and star formation rate (SFR) of our Galaxy?  The BMM methodology is a vetted and powerful tool for answering these questions (see, e.g., \citealp{Press97,LangHogg}), which enabled us to produce results that account for the possibility that any single measurement is potentially flawed or suffers from systematic errors, given the complexities of modeling the MW (we refer the reader to that work for more details). The same paper describes the Monte Carlo techniques we have used for determining the total stellar mass of the MW by incorporating a model for the stellar disk.  Initially, we utilized the parameters of the dynamical mass model from \citet{BovyRix13}, but in this paper we utilize the updated model from LN16 (q.v. below).

In L15 we demonstrated that identifying a sample of MW-analog galaxies from SDSS data enabled us to convert our posterior knowledge of the Galactic M$_\star$ and SFR from LN15 into tight constraints on the global photometric properties (i.e., color and luminosity) of the MW as the SDSS instrument would measure them face-on from across cosmic distances. In Table \ref{table:mw}, we list absolute magnitudes for the MW calibrated to the same scale as those observed for other spiral galaxies, for both $ugriz$ and $UBVRI$ filter passbands.

In LN16 we have applied the BMM methodology from LN15 to nearly 30 different estimates of the scale length of the MW disk in the literature based on observations of either visible or infrared starlight from the Galaxy. In Table \ref{table:mw}, our estimate of $R_{\rm d} = 2.71^{+0.22}_{-0.20}$ kpc is the result from considering visible light only, which corresponds most closely with the extragalactic scale lengths measured by H12 in the $i$-band. In comparison, we found the disk scale length to be $2.51^{+0.15}_{-0.14}$ kpc when including only those measurements that rely on infrared starlight, differing by only 0.8$\sigma$.

Given that infrared light is expected to trace more closely the distribution of stellar mass, in LN16 we next used this value to produce an updated model of the stellar disk which also accounts for local stellar density variations due to spiral structure and its associated uncertainty --- a correction that typically has been ignored in MW models \citep{Hessman15}.  We then combined this with our bulge+bar mass estimate from LN15 to determine a total stellar mass for the MW of $5.67^{+1.53}_{-1.11}\times10^{10}$ M$_\odot$, or equivalently $\log({\rm M}_\star/{\rm M}_\odot) = 10.75\pm0.10$.  This represents only a slight change from our original estimate of $10.78 \pm 0.08$ from LN15 when using the dynamical model of the disk from \citet{BovyRix13}. The absolute magnitudes listed in Table \ref{table:mw} are derived using this revised M$_\star$ estimate, together with the unchanged SFR value; these differ only slightly from the original values in L15.

We note that all estimates of Galactic properties described so far consistently utilize a prior on the Galactocentric radius of the Sun of $R_0=8.33\pm0.35$ kpc, based on the work of \citet{Gillessen09}. This value stems from careful consideration of both statistical and systematic uncertainties, and represents an up-to-date, geometric estimate that is still consistent with essentially all recent $R_0$ measurements, which predominantly fall in the range of $\sim$8--8.5 kpc (see, e.g., \citealp{Schonrich12}, \citealp{Do13}, \citealp{Reid14}, and \citealp{Chatzopoulos15} for recent estimates that closely match our prior). Most importantly, error bars quoted for all MW quantities incorporate the uncertainties associated with our limited knowledge of $R_0$. We will demonstrate in \S\ref{sec:discussion_prior_tfr} that prior attempts at discussing the MW in the context of the TFR were problematic due, in large part, to their treatment of $R_0$.

We employ the International Astronomical Union standard value of 220 km s$^{-1}$ for the rotation speed of the Galactic disk \citep{KLB86} and ascribe a conservative 10\% error estimate (in line with preceding studies). We find that this agrees well with the current set of estimates available in the literature (note that these typically scale proportionately with $R_0$ due to the conversion from angular speed to tangential speed; see the description of estimates in \citealp{Majewski08,Reid14}, as well as those of \citealp{BovyRix13,Sirko04,Piffl14}). Several of the more recent measurements have yielded values closer to $\sim$240--250 km s$^{-1}$.  We note here that if we were to increase our adopted value of $V_{\rm rot}$ by 1$\sigma$ (i.e., $+0.04$ dex in $\log V_{\rm rot}$) to better match these, tension between the MW and the TFR increases moderately, but we still find the MW to be consistent with the relation, given its scatter. Furthermore, tension with the $LVR$ relation also increases, and our overall conclusions would be unchanged.	

Estimates of the neutral hydrogen gas mass (M$_{\rm HI}$) of the MW are sparse in the literature. Following what \citet{Flynn} have done (but omitting molecular hydrogen to match our extragalactic data), we can scale the HI estimates tabulated by \citet{Dame93} to our choice of $R_0 = 8.33$ kpc in order to find estimates of ${\rm M}_{\rm HI} = 3.3\times10^9$ M$_\odot$ based on the work of \citet{Henderson82}, or ${\rm M}_{\rm HI} = 6.7\times10^9$ M$_\odot$ by combining values for inside and outside the solar circle from \citet{Liszt92} and \citet{Wouterloot90}. More recently, \citet{Kalberla07} have made use of much more sensitive Galactic HI line surveys with improved coverage, both spatially and kinematically, in order to produce a model of the interstellar medium (ISM) that yields a total HI mass of $8\times10^9$ M$_\odot$ \citep{KalberlaKerp09}. When scaled to our choice of $R_0$ this becomes ${\rm M}_{\rm HI} = 7.7\times10^9$ M$_\odot$. Lastly, the \citet{BovyRix13} model of the Galactic disk assumes a smaller total mass for the ISM of $7\times10^9$ M$_\odot$ (insensitive to the choice of $R_0$). If we apply here the HI-to-ISM mass ratio ($=0.65$) from the \citet{KalberlaKerp09} model, this yields an estimate of ${\rm M}_{\rm HI} = 4.5\times10^9$ M$_\odot$. Therefore, we take the mean and standard deviation of these four values in order to produce an aggregate estimate of ${\rm M}_{\rm HI} = 5.55\pm1.74\times10^9$ M$_\odot$, which we adopt for this study. Finally, we calculate the baryonic mass of the MW as ${\rm M}_{\rm bar} = {\rm M}_\star + 1.4 {\rm M}_{\rm HI} = 6.45^{+1.54}_{-1.14}\times10^{10}$ M$_\odot$, or equivalently $\log({\rm M}_{\rm bar}/{\rm M}_\odot) = 10.81\pm0.09$.

\section{Investigating Tully-Fisher Relations} \label{sec:tfr} 

We begin by first investigating how the MW fits with the TFR.  We explore seven versions of the TFR by examining the relationship between rotation speed and a variety of quantities related to luminosity: rest-frame absolute magnitudes ($^0\!M$) in all five SDSS $ugriz$ passbands, log total stellar masses ($\log {\rm M}_\star$), and log baryonic masses ($\log {\rm M}_{\rm bar}$). As mentioned above, we measure each relation from our sample of 258 spiral galaxies observed by the SDSS that display moderate inclinations, allowing all quantities of interest for this study to be measured cleanly (cf. \S\ref{sec:data_SDSS}).  

\subsection{Methods} \label{sec:methods_tfr} 

In each case, we perform a least-squares optimization to fit the inverse relation, where we treat $\log V_{\rm rot}$ as the dependent variable; this is common practice to avoid any sample-selection biases in modeling the data that would be tied to errors in measurements of $L$ \citep{Schechter80,Tully88}. Generically, we fit for $\log V_{\rm rot} = a + b\log L$, where $a$ is the zero point, $b$ is the slope, and $\log L$ can be any of the seven proxies listed above. We utilize the {\tt scipy.optimize.curve\_fit} package to perform a non-linear least-squares fit of each model to the data, providing the $a$ and $b$ values of the optimal relation. We then determine the scatter ($\sigma_{\rm TF}$) about the relation by measuring the standard deviation of the $\log V_{\rm rot}$ values about the best-fit line (typically $\sim$0.08 dex in $\log V_{\rm rot}$). Lastly, we measure the offset of the MW data point from the line along the $\log V_{\rm rot}$ direction using the combination in quadrature of both $\sigma_{\rm TF}$ and MW errors. 

In Table 2 we tabulate values pertinent to each TFR comparison, including $a$, $b$, $\sigma_{\rm TF}$, and the level of consistency of the MW data point. Here, we also show analogous comparisons to TFRs found using the spiral galaxy sample with no inclination cuts applied, as well as to TFRs found by other authors.

We have performed several tests on the robustness of our results.  For example, we have conducted a ``forward'' fitting of each TFR to our extragalactic data, i.e., where we model $\log L = a^\prime + b^\prime \log V_{\rm rot}$, equivalent to assuming that Malmquist bias \citep{Teerikorpi97} is of minimal consequence.  As is typically observed, the forward TFRs are shallower than their corresponding inverse relations, but with larger zero points (or intercepts), and have similar scatters \citep[cf.][]{Pizagno07}.  Nonetheless, in all cases we find that the intersection of the forward and inverse relations occurs very near the MW's $L$ and $V_{\rm rot}$; our results would remain the same regardless of how we fit for the relation.  We also have tested for any bias in our fits due to large outliers in the extragalactic data.  Here, in place of our standard least-square optimization, we employ the {\tt Statsmodels} Python module to perform
\begin{figure*}[h!]
\centering
\vspace{.3in}
\includegraphics[page=1, trim= .7in 1.7in .68in 1in, clip=true, width=\textwidth]{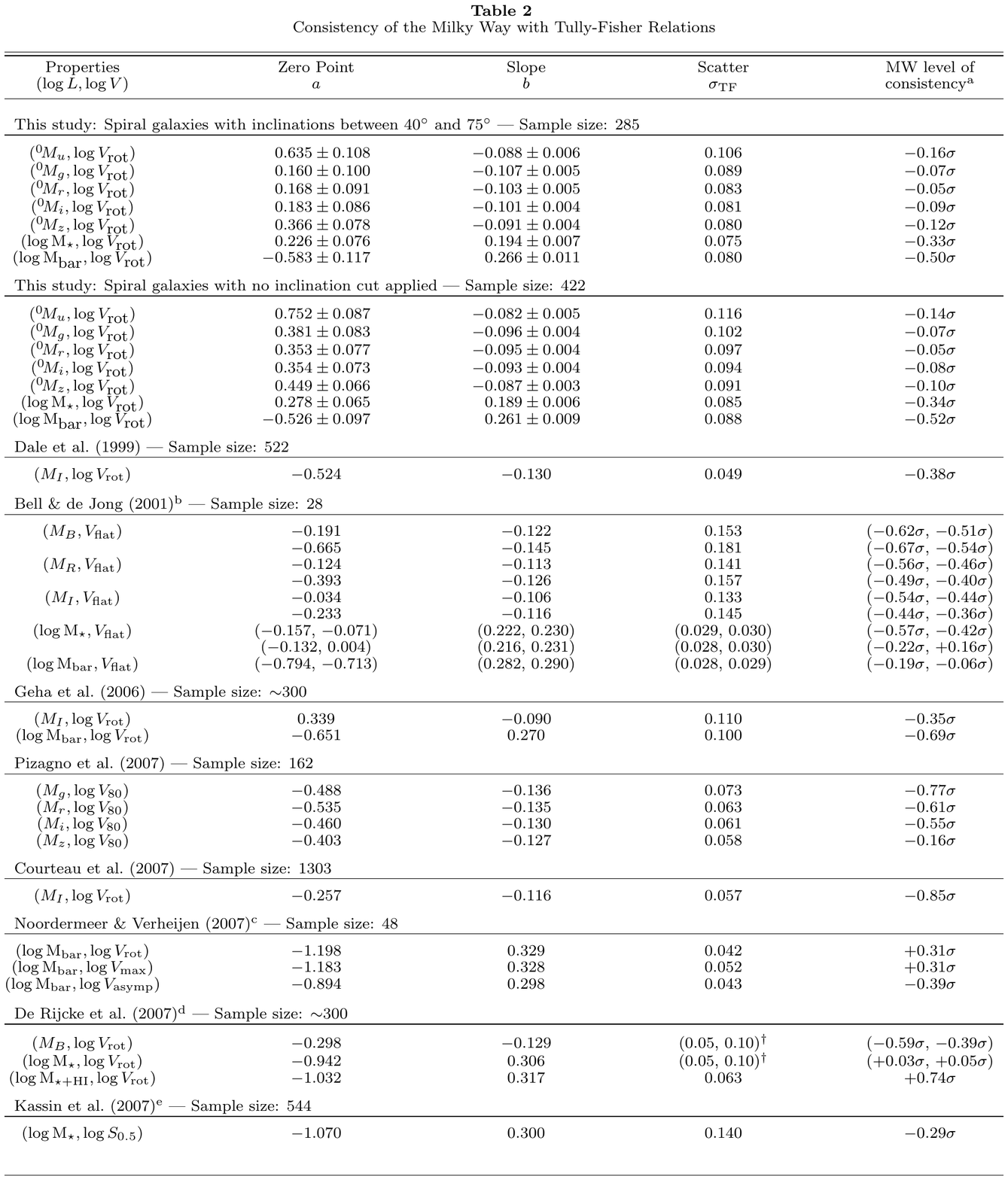}
\end{figure*}
\begin{figure*}[h!]
\centering
\vspace{.3in}
\includegraphics[page=2, trim= .7in 0in .68in 1.125in, clip=true, width=\textwidth]{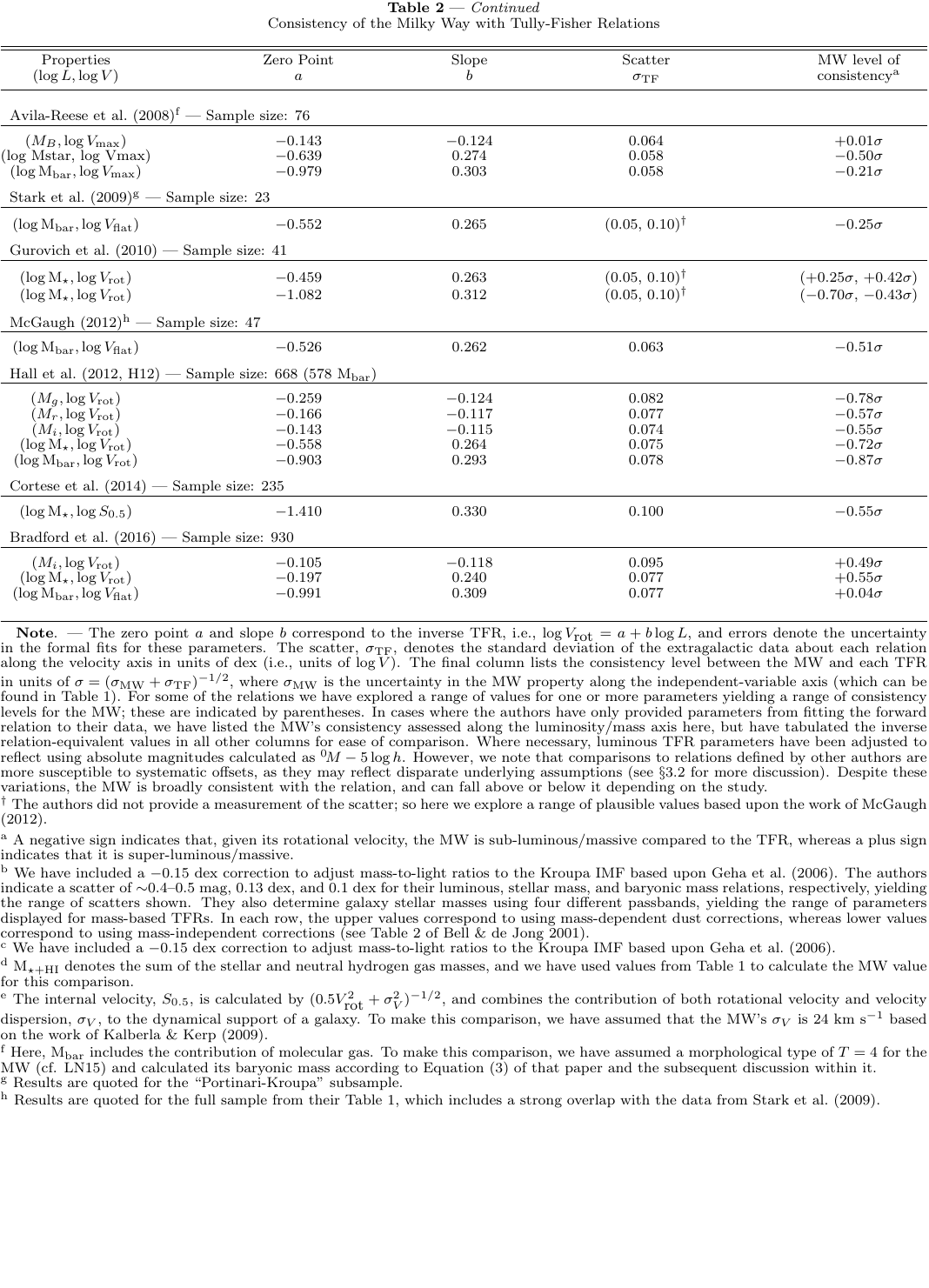}
\end{figure*}
\clearpage
\noindent robust linear modeling using M-estimators \citep{Maronna06}.  In particular, we have used both Huber's and Tukey's estimator functions (with tuning constants of 1.345$\sigma_{\rm MAD}$ and 4.685$\sigma_{\rm MAD}$, respectively, where $\sigma_{\rm MAD}$ is the median absolute deviation measured for the entire sample) to deweight objects with large residuals in the forward or inverse fits of the TFR.  In all cases, this yields negligible changes in the best-fit line, and, again, our conclusions are negligibly affected.

\subsection{Results} \label{sec:results_tfr} 

\begin{figure*}
\centering
\includegraphics[trim=0.1in 0.1in 0.1in 0.1in, clip=true, width=\textwidth]{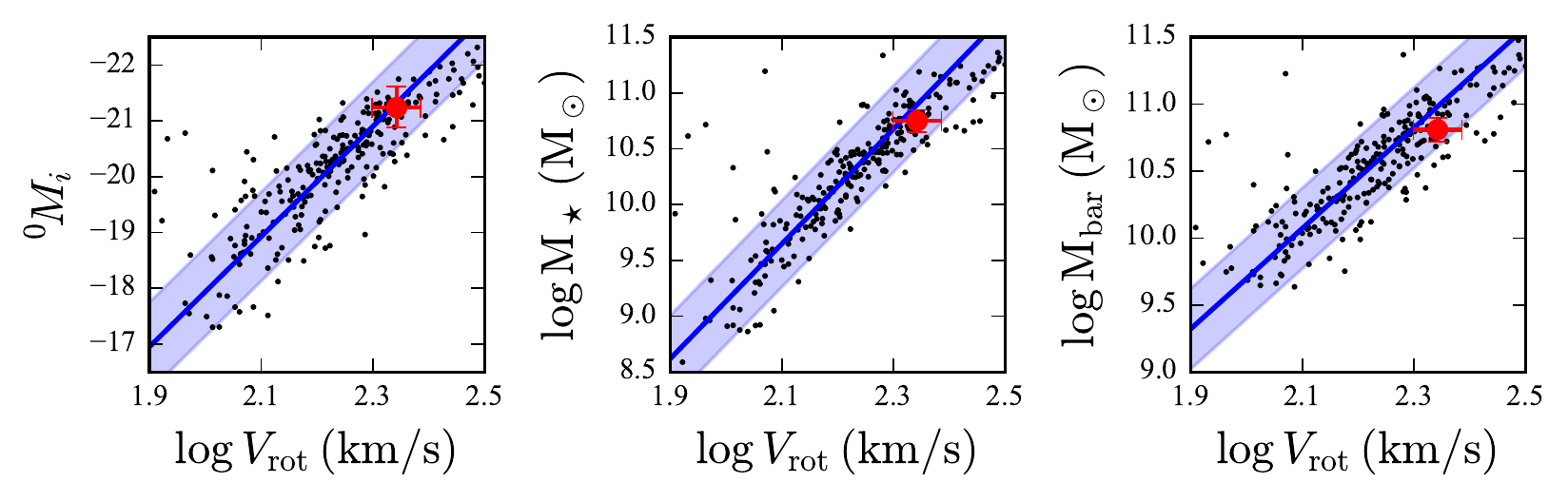}
\caption{\footnotesize Tully-Fisher relations (TFRs) defined by the rest-frame absolute $i$-band magnitudes ($^0\!M_i$; left panel), total stellar masses ($\text{M}_\star$; center panel), and baryonic masses ($\text{M}_\text{bar}$; right panel) of our sample of 258 SDSS spiral galaxies with moderate inclinations, shown as black dots.  In each panel, these properties are plotted as a function of the disk rotational velocity ($V_\text{rot}$).  The blue line and shaded blue region shows our best-fit (inverse) TFR and 1$\sigma$ range, respectively, measured from the black dots (see \S\ref{sec:results_tfr}).  Lastly, in each panel we overlay the MW datapoint with error bars in red, corresponding to the values in Table \ref{table:mw}.  In all cases, the properties of the MW are in excellent agreement with the TFR (see Table 2).}
\label{fig:tfr}
\end{figure*}

Based on our results listed in Table 2, we find that, in all cases, the properties of the MW are in excellent agreement with the TFR. For those relations measured in this study, where Galactic and extragalactic measurements are ensured to be on equal footings and when using SDSS magnitudes, the MW is consistent with all TFRs with deviations of less than 0.20$\sigma$. When using total stellar mass or baryonic mass, we find the MW data point to be consistent with the relation at the 0.33$\sigma$ and 0.50$\sigma$ levels, respectively. Figure \ref{fig:tfr} illustrates the analyses for $i$-band absolute magnitude ($^0\!M_i$), total stellar mass (M$_\star$), and baryonic mass (M$_{\rm bar}$).  Table 2 illustrates that these results are robust to using our spiral galaxy sample with or without the inclination cut applied.

MW properties appear to be in good overall agreement with TFRs measured in other studies as well, with consistencies that generically fall well below the 1$\sigma$ threshold.  It is important to note that comparisons to relations defined by other authors are more susceptible to systematic offsets, as they can differ either because they rely on quantities calculated differently from those used here or because they employ systematically different samples (see \citealp{Bradford16} for a thorough investigation of these effects).  Despite these variations, the MW is broadly consistent with the relation, and can fall above or below it depending on the study (denoted by the +/$-$ in the final column of Table 2).

\section{Investigating 3-dimensional $LVR$ Relations} \label{sec:lvr} 

As mentioned above, measures of galaxy size, or radius ($R$), also obey scaling relations with $L$ and $V_{\rm rot}$. The distribution of starlight from a spiral galaxy's disk component can be fit well to first order by an exponentially declining profile (cf. LN16). We denote the radius where light has dimmed by a factor of 1/$e$ (or $\sim$37\%) compared to the brightness at the center of the disk as the radial scale length ($R_{\rm d}$), which serves as a standardized measure of size for spiral galaxies. We next compare the MW's properties to scaling relations found in 3-dimensional luminosity-velocity-radius ($LVR$) parameter spaces.

\subsection{Methods} \label{sec:methods_lvr} 

To investigate scaling relations that include galaxy size, we now add a third axis, corresponding to disk scale length measurements, to each of the seven TFR diagrams investigated above. In each 3D parameter space, we perform principal component analysis \citep[PCA;][]{Jolliffe02} on the extragalactic data to determine the best-fit $LVR$ relation using the {\tt numpy.linalg.svd} Python routine. We use this function to factor the $LVR$ data in matrix form via singular value decomposition in order to determine three orthonormal eigenvectors, or principal components. The first principal component (PC1) indicates the direction in $LVR$ space for which the data displays maximal variance. The second principal component (PC2) indicates the direction of next largest variance that is perpendicular to PC1, and the third principal component (PC3) is the direction orthogonal to both the PC1 and PC2. We define each best-fit $LVR$ relation as the line parallel to PC1 that passes through the mean position of the data.  Next, in the PC2-PC3 plane (i.e., the plane orthogonal to the PC1), we determine the 68\% and 95\% scatter about the relation by finding the most compact elliptical contours that envelope those percentages of the data.  Figure \ref{fig:PC} shows this analysis for the particular case of using $^0\!M_i$ for luminosity, where we show each principal component in $LVR$ space and also show the data projected onto each of the coordinate planes in PCA space. We subtract off the mean and renormalize the data into units of the standard deviation along each of the $LVR$ axes in order to produce the PCA-plane projections.

To quantitatively assess the MW's consistency with each $LVR$ relation, we perform a series of Monte Carlo simulations where we randomly generate MW data points by drawing values from the probability distribution functions (PDFs) for Galactic properties listed in Table \ref{table:mw}; simultaneously we bootstrap resample from our spiral galaxy data. For each realization, this involves the following:
\begin{enumerate}
\item Generate a MW data point by randomly drawing values for $\log L$, $\log V_{\rm rot}$, and $\log R_{\rm d}$ independently of each other from the PDFs described in Table \ref{table:mw};
\item Randomly draw galaxies from our extragalactic sample, each with equal probability and with repeats allowed, until the original sample size is achieved;
\item Perform a new PCA on this bootstrapped sample;
\item Measure the distance of all data points from the PC1 by calculating their elliptical radii from the origin in the PC2-PC3 plane by $\left[\left(d_{\rm PC2}/\sigma_{\rm PC2}\right)^2 + \left(d_{\rm PC3}/\sigma_{\rm PC3}\right)^2\right]^{1/2}$; and,
\item Finally, calculate the fraction of objects that lie inside of the ellipse passing through the MW data point, which we denote as $f_{\rm <MW}$.
\end{enumerate}

We perform this analysis $10^4$ times, resulting in a distribution of $f_{\rm <MW}$ values that incorporates all uncertainties in MW properties and all underlying uncertainties in fitting the $LVR$ relation. For each relation that we investigate, we list in Table \ref{table:LVR} the mean location ($\vec{\mu}$) of the data, the standard deviation ($\vec{\sigma}$) along each $LVR$ axis, the PC1 eigenvector, and $f_{\rm <MW}$.

Lastly, independent of the PCA described above, we also investigate the disk scale length that the $LVR$ data would predict for a galaxy of the MW's luminosity and rotational velocity. Here, we perform multiple linear regression on the extragalactic data, modeling $\log R_{\rm d}$ as a function of the predictor variables $\log L$ and $\log V_{\rm rot}$. More explicitly, we can write
\begin{equation} \label{eq:1}
\log R_{\rm d} = \alpha \log L + \beta \log V_{\rm rot} + \gamma + \epsilon,
\end{equation}
where $\alpha$, $\beta$, and $\gamma$ are the fitted coefficients and $\epsilon$ reflects the residual between a given $\log R_{\rm d}$ value and the prediction from that galaxy's $\log L$ and $\log V_{\rm rot}$. By inspecting a normal probability plot for $\epsilon$ using the {\tt scipy.stats.probplot} Python routine, we find that the residual in all cases is well approximated by a Gaussian distribution. We then perform Monte Carlo simulations where we randomly draw values independently from the MW probability distributions for $\log L$ and $\log V_{\rm rot}$ listed in Table \ref{table:mw} and from the distribution of $\epsilon$ values; each set of values $(\log L, \log V_{\rm rot}, \epsilon)$ is then combined with the best-fit coefficients in Equation (1) in order to build up a distribution of $\log R_{\rm d}$ values. The median value and 68\% confidence interval for $R_{\rm d}$ measured from each distribution is listed in the final column of Table \ref{table:LVR}. These correspond to the range of disk scale lengths that one would expect to obtain from an externally measured SDSS $i$-band image of a galaxy with the MW's $L$ and $V_{\rm rot}$.

One might be concerned that these predictions could be affected by attenuation bias; i.e., the reduction of the amplitude of measured regression coefficients from their underlying values due to errors in the predictor variables. We test for this by adding random Gaussian noise to the predictors, drawn from a normal distribution of mean 0 and standard deviation $\sigma^\ast$, where $\sigma^\ast$ reflects the typical measurement error (see \S4.1 of H12). We find that this produces a $<$0.1\% shift in the predicted $\log R_{\rm d}$, indicating that our results are robust to attenuation bias.

\subsection{Results} \label{sec:results_lvr}
\begin{figure*}
\centering
\includegraphics[trim=0in 0.13in 0in 0.1in, clip=true, width=\textwidth]{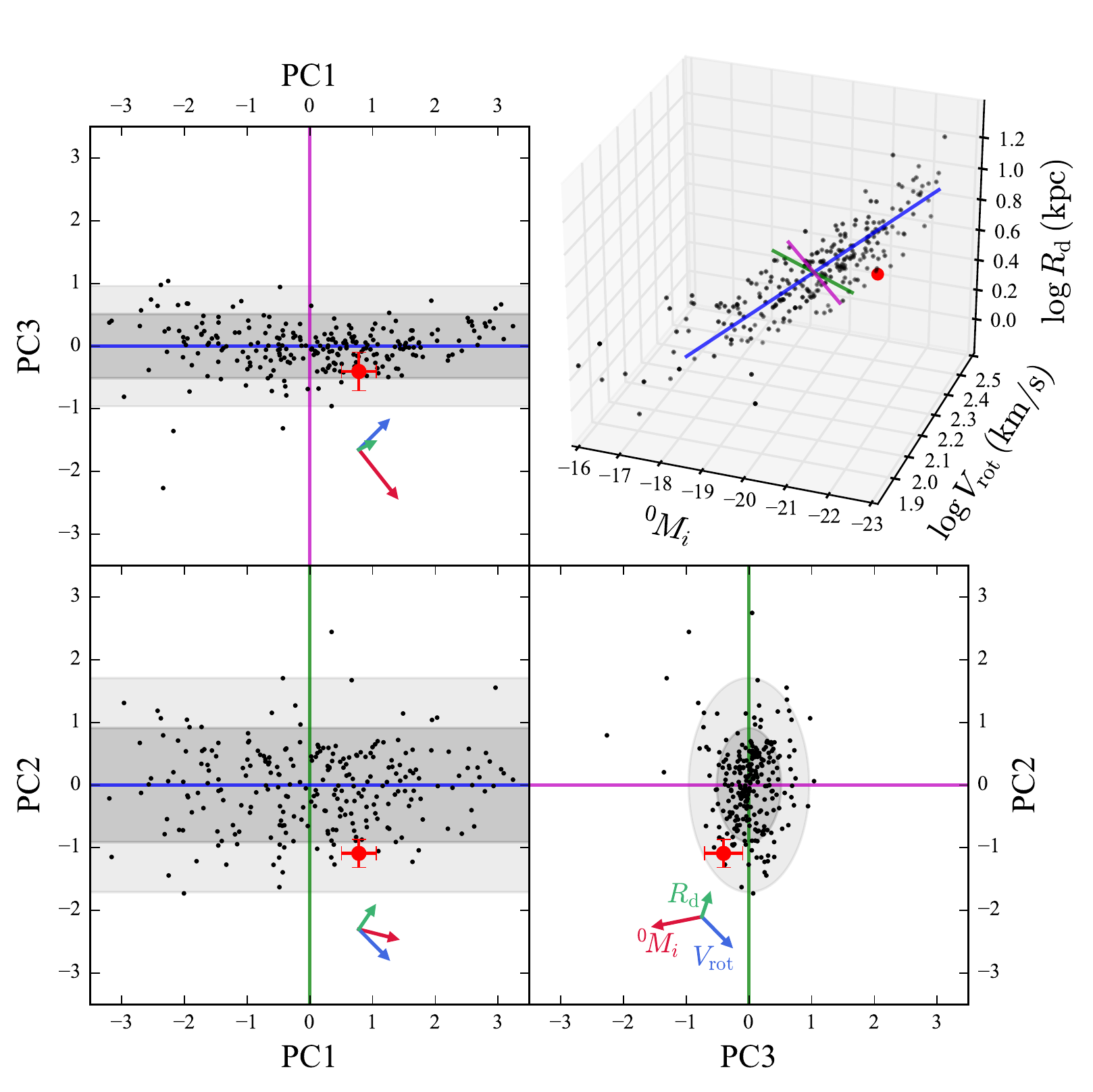}
\caption{\footnotesize The distribution of our SDSS spiral galaxy sample in luminosity-velocity-radius ($LVR$) space (top right) and PCA space (all other panels).  In the top right panel, we show the 3D distribution using rest-frame absolute $i$-band magnitude, $^0\!M_i$, log of the disk rotational velocity, $\log V_\text{rot}$, and log of the disk scale length, $\log R_{\rm d}$.  The direction of the first, second, and third principal components (PC1, PC2 and PC3; see \S\ref{sec:methods_lvr}) describing the black data are depicted by blue, green, and magenta lines, respectively.  We overlay the MW datapoint as a red dot.  The other three panels show the data projected onto each of the coordinate planes in PCA space after renormalizing to units of standard deviation along each of the $LVR$ axes and subtracting off the mean, marked by the intersection the PCA axes.  The 68\% and 95\% confidence regions are shaded in dark and light gray, respectively.  We overlay the MW datapoint in red with error bars.  The bottom right panel shows the projection of the data onto the PC2-PC3 plane --- i.e., the plane orthogonal to the blue line, which we take as our best-fit $LVR$ relation; in this space we have determined that the MW lies just inside of the 95\% confidence region for the 3D relation (see \S\ref{sec:methods_lvr} for more details).  Lastly, we illustrate that this discrepancy is predominantly due to the small size of the Galactic disk by showing the effect on the MW datapoint due to a 3$\sigma$ shift in $^0\!M_i$ (red arrow), $V_\text{rot}$ (blue arrow), and $R_{\rm d}$ (green arrow).  Large changes in \emph{both} $^0\!M_i$ and $V_\text{rot}$ would need to conspire together to reduce this tension, whereas increasing values of $R_{\rm d}$ would pushes it directly toward the $LVR$ relation, in fact landing nearly on it when reaching a value of $\sim$5.25 kpc (corresponding to a $\sim$11.5$\sigma$ increase; see \S\ref{sec:lvr} for a more rigorous assessment of this offset).}
\label{fig:PC}
\end{figure*}

Including all known sources of uncertainty, we find that the fraction of objects lying closer to the $LVR$ relation than the MW datapoint is $\sim$0.90 (with minor variations depending on the proxy used for $L$). Hence, our Galaxy lies just inside of the 95\% confidence region boundary for the $LVR$ relation, but well outside of the 68\% confidence region. We illustrate the $LVR$ analysis using $i$-band absolute magnitude in Figure \ref{fig:PC}.  In all cases, we find that the offset of the MW data point is consistent with being entirely along the direction of the scale length axis. Considering only measurement errors, this constitutes $\sim$9.5$\sigma$ evidence that the MW is unusually compact compared to typical spiral galaxies of its $L$ and $V_{\rm rot}$. However, $\sim$10\% of galaxies lie even further from the $LVR$ relation; so the MW is clearly undersized, but not extraordinarily so.

From our exercise of turning this analysis around and performing multiple linear regression on our extragalactic sample, and in particular to predict $R_{\rm d}$ values from a galaxy's $^0\!M_i$ and $V_{\rm rot}$, we find that the expected scale length of the MW disk, as measured externally from photometric techniques (such as in H12), would be $4.92^{+1.82}_{-1.53}$ kpc. This is nearly double the observed value that we have adopted here of $2.71^{+0.22}_{-0.20}$, and these estimates are inconsistent with each other at the $\sim$1.4$\sigma$ confidence level (incorporating both MW errors and the observed scatter about the regression relation).

\section{Discussion} \label{sec:discussion}

\subsection{Comparison to Prior Investigations of the Milky Way's Consistency with the TFR} \label{sec:discussion_prior_tfr}

Our finding that the MW's properties are highly consistent with the TFR contrasts with prior investigations that labeled our Galaxy a 1--1.5$\sigma$ outlier to the relation. It is important to note that all of these studies, including this one, adopted the same value of 220 km s$^{-1}$ with a $\sim$10\% uncertainty for the MW's $V_{\rm rot}$. We conclude instead that the difference is a result of systematic differences in Galactic luminosity estimates arising primarily, if not entirely, from varying choices of Galactic parameters. Specifically, these prior studies adopted luminosity estimates derived from MW models that depend strongly on the assumed values of $R_0$ and $R_{\rm d}$. It is a common practice in Galactic astronomy to adopt a fixed, consensus value for $R_0$ --- typically 8 or 8.5 kpc with no uncertainties included --- so that one can straightforwardly compare results between various MW models. This becomes less sensible, however, once one becomes interested in Galactic-to-extragalactic comparisons (as in this study), as any error in the assumed value of $R_0$ will become a source of systematic uncertainty.

For example, \citet{Flynn} analyzed data from the Hipparcos and Tycho stellar surveys, determining the local surface luminosity in the $I$-band to be 29.54 $L_\odot$ pc$^{-2}$. They then extrapolated this to a global luminosity estimate using an exponential model of the disk by assuming $R_0 = 8$ kpc (with no uncertainty) and exploring values of $R_{\rm d}$ in the range of 2.5--5.0 kpc. After including a 10\% correction for spiral arms and adopting a near-infrared bulge luminosity of $\sim$10$^{10}$ $L_\odot$ (ignoring any uncertainty), they find a global value of $M_I = -22.3\pm0.2$. Using the same value of $V_{\rm rot}$ for the MW that we have adopted for this study, they then found that our Galaxy is under-luminous by $\sim$1$\sigma$ with respect to the $I$-band TFR defined by external spirals from \citet[][though see \citealp{BovyRix13} for an update]{Dale99}. We find that if the \citet{Flynn} estimates were to instead utilize the same values of $R_0$ and $R_{\rm d}$ as we have employed for this study, they would yield $M_I = -22.73\pm0.25$, which is in excellent agreement with our slightly dimmer estimate of $^0\!M_I - 5 \log(h/0.7) = -22.61^{+0.36}_{-0.39}$ from L15. Consequentially, the MW would appear only $\sim$0.05 mag dimmer than the $I$-band TFR from \citet{Dale99} and would be consistent with it at the 0.1$\sigma$ level.

Similarly, \citet{Hammer07} used a $K$-band absolute magnitude estimate of $-22.15$ (converted to the AB system) for the MW from \citet{DrimmelSpergel01}, based on an exponential disk model with $R_0 = 8$ kpc (with no uncertainty) and $R_{\rm d} = 2.26\pm0.08$ kpc, as well as the same value of Galactic $V_{\rm rot}$ adopted here, to find that our Galaxy is again $\sim$1$\sigma$ under-luminous with respect to the $K$-band TFR defined by combining three different local galaxy samples. Unfortunately, the MW model of \citet{DrimmelSpergel01} is complex, with 19 free parameters in total, making it difficult to straightforwardly scale to different choices of Galactic parameters. However, if we use the standard exponential model in order to approximate a first-order correction, we find that translating to the values of $R_0$ and $R_{\rm d}$ that we have adopted corresponds to an increase of approximately $-0.3$ mag for $M_K$, which would leave the MW $0.3\pm0.5$ mag dimmer than the $K$-band TFR, and hence consistent with it at the 0.6$\sigma$ level.

On the other hand, \citet{Malhotra96} adopted $R_0 = 8.5$ kpc and the same $V_{\rm rot}$ employed here to derive $M_K = -24.06$ for the MW using a model similar to that of \citet{DrimmelSpergel01}. This indicated our Galaxy to be $\sim$1$\sigma$ \emph{over}-luminous with respect to the $K$-band TFR relation that they find for five nearby well-measured galaxies with Cepheid-based distances. Again corrections are difficult to make and the sample size for this TFR is small, but reducing the value of $R_0$ to match current data and incorporating the associated uncertainties would yield a dimmer $M_K$ value and eliminate much of this tension, just as in the other cases.

A related analysis with broadly similar conclusions was recently published in \citet{McGaugh16}.  These authors utilized a novel approach to determining Galactic properties: they iteratively adapted mass models of the MW by hand until they produce a terminal velocity curve that by-eye matches HI and CO observations taken in the first and fourth quadrant of the Galaxy, with a particular focus on reproducing the ``bumps and wiggles'' in that data. The authors suggest that this should account for substructure (e.g., spiral arm) features in the stellar surface density that would be unaccounted for by assuming a smooth exponential profile \citep[cf.][]{Sancisi04}. They explore six MW models in total, corresponding to a range of parameter values, specifically: ${\rm M}_\star = 5$--$6\times10^{10}$ M$_\odot$, $V_{\rm rot} = 222$--$233$ km s$^{-1}$, and $R_{\rm d} = 2.0$--2.9 kpc; all of these values assume $R_0 = 8$ kpc (with no uncertainty) and the authors claim that varying this assumption would affect their results, particularly $V_{\rm rot}$ and M$_\star$ estimates, in a non-straightforward way. Nevertheless, these values are all in excellent agreement with the ones we have used in this study. They also adopt a total gas mass (atomic + molecular, corrected for He and metals) of $1.18\times10^{10}$ M$_\odot$ (with no uncertainty ascribed to this value) which they add to M$_\star$ to yield ${\rm M}_{\rm bar} = 6.18$--$7.18\times10^{10}$ M$_\odot$.  Finally, they compare their MW results to extragalactic data (a subset of which is used in this study), assessing qualitatively that the MW appears normal with respect to the baryonic TFR, but is somewhat compact for its mass. While there are significant differences between the assumptions and methodologies underlying the Galactic versus extragalactic data that are compared within \citet{McGaugh16} --- differences that we have sought to minimize in this work --- the results from \citet{McGaugh16} nevertheless broadly agree with the conclusions presented here.
\begin{deluxetable*}{cccccc}
\tablenum{3}\label{table:LVR}
\tablewidth{\textwidth}
\tablecaption{Consistency of the Milky Way with $LVR$ Relations}
\tablehead{$\log L$ & Sample $\vec{\mu}$ & Sample $\vec{\sigma}$ & PC1 & $f_{<\text{MW}}$ & Predicted \\
 proxy & & & eigenvector & & MW $R_{\rm d}$ \\
 (1) & (2) & (3) & (4) & (5) & (6)}
\startdata
\\[-0.5ex]
\multicolumn{5}{l}{Spiral galaxies with $40^{\circ}<i<75^{\circ}$ -- Sample size: 285} \\
\hline \\[-1ex]
$^0\!M_u$ & $(-17.963,2.219,0.547)$ & $(1.106,0.144,0.205)$ & $(-0.592,0.583,0.556)$ & $0.88^{+0.07}_{-0.13}$ & $4.75^{+1.98}_{-1.51}$ \\
$^0\!M_g$ & $(-19.242,2.219,0.547)$ & $(1.064,0.144,0.205)$ & $(-0.610,0.572,0.548)$ & $0.90^{+0.06}_{-0.13}$ & $4.97^{+1.92}_{-1.54}$ \\
$^0\!M_r$ & $(-19.825,2.219,0.547)$ & $(1.139,0.144,0.205)$ & $(-0.612,0.575,0.543)$ & $0.90^{+0.06}_{-0.11}$ & $4.99^{+1.90}_{-1.53}$ \\
$^0\!M_i$ & $(-20.106,2.219,0.547)$ & $(1.179,0.144,0.205)$ & $(-0.613,0.577,0.540)$ & $0.89^{+0.07}_{-0.12}$ & $4.92^{+1.82}_{-1.53}$ \\
$^0\!M_z$ & $(-20.299,2.219,0.547)$ & $(1.314,0.144,0.205)$ & $(-0.610,0.586,0.533)$ & $0.84^{+0.08}_{-0.12}$ & $4.62^{+1.81}_{-1.49}$ \\
$\log \textrm{M}_\star$ & $(10.262,2.219,0.547)$ & $(0.634,0.144,0.205)$ & $(0.612,0.589,0.528)$ & $0.82^{+0.07}_{-0.11}$ & $4.53^{+1.67}_{-1.47}$ \\
$\log \textrm{M}_\textrm{bar}$ & $(10.520,2.219,0.547)$ & $(0.450,0.144,0.205)$ & $(0.615,0.570,0.545)$ & $0.84^{+0.08}_{-0.13}$ & $4.42^{+1.39}_{-1.25}$ \\[1.5ex]

\multicolumn{5}{l}{Spiral galaxies with no inclination cut -- Sample size: 422} \\
\hline \\[-1ex]
$^0\!M_u$ & $(-17.732,2.209,0.565)$ & $(1.154,0.150,0.214)$ & $(-0.583,0.584,0.565)$ & $0.91^{+0.05}_{-0.10}$ & $5.09^{+2.50}_{-1.65}$ \\
$^0\!M_g$ & $(-19.015,2.209,0.565)$ & $(1.139,0.150,0.214)$ & $(-0.599,0.577,0.555)$ & $0.91^{+0.06}_{-0.08}$ & $5.24^{+2.37}_{-1.61}$ \\
$^0\!M_r$ & $(-19.611,2.209,0.565)$ & $(1.210,0.150,0.214)$ & $(-0.603,0.578,0.550)$ & $0.92^{+0.05}_{-0.09}$ & $5.27^{+2.43}_{-1.59}$ \\
$^0\!M_i$ & $(-19.896,2.209,0.565)$ & $(1.253,0.150,0.214)$ & $(-0.604,0.579,0.547)$ & $0.91^{+0.05}_{-0.09}$ & $5.23^{+2.27}_{-1.66}$ \\
$^0\!M_z$ & $(-20.120,2.209,0.565)$ & $(1.357,0.150,0.214)$ & $(-0.604,0.583,0.543)$ & $0.89^{+0.05}_{-0.09}$ & $4.92^{+2.11}_{-1.49}$ \\
$\log \textrm{M}_\star$ & $(10.212,2.209,0.565)$ & $(0.654,0.150,0.214)$ & $(0.607,0.585,0.537)$ & $0.86^{+0.06}_{-0.09}$ & $4.85^{+1.97}_{-1.53}$ \\
$\log \textrm{M}_\textrm{bar}$ & $(10.493,2.209,0.565)$ & $(0.465,0.150,0.214)$ & $(0.611,0.567,0.552)$ & $0.89^{+0.06}_{-0.09}$ & $4.68^{+1.61}_{-1.25}$
\enddata
\tablecomments{Column (1) shows the proxy used for $\log L$ that is combined with $\log V_{\rm rot}$ and $\log R_{\rm d}$ to construct each $LVR$ relation. Values in Columns (2)--(4) are measured from our nominal spiral galaxy sample. Values in Column (5) are produced from PCA using the Monte Carlo techniques described in \S\ref{sec:methods_lvr}. These values reflect the 16th, 50th, and 84th percentiles of the resulting distribution of $f_{<\text{MW}}$; i.e., the fraction of the bootstrapped sample closer to its first principal component than the current realization of the MW datapoint. Values in Column (6) represent the $\log R_{\rm d}$ predicted for the MW in units of kpc based on its $L$ and $V_{\rm rot}$, as described in \S\ref{sec:methods_lvr}.}
\end{deluxetable*}

\subsection{The Emergent Picture of a ``Too-Small'' Milky Way Galaxy} \label{sec:discussion_toosmall}

As we have detailed in this paper, comparing measurements of the MW's luminosity/mass, rotational velocity, and disk scale length to both Tully-Fisher and $LVR$ relations for other spiral galaxies indicates that our Galaxy is approximately two times smaller (more compact) than is typical for its $L$ and $V_{\rm rot}$. This is potentially valuable knowledge for determining what evolutionary histories are possible for our Galaxy, especially as simulations of disk galaxies with an assortment of merger histories are growing in both number and mass resolution \citep[see, e.g.,][]{Martig14,Taylor15}.  Therefore, it is important to know what potential concerns and additional support can be associated with our results, and we discuss these here.

\subsubsection{Potential Concerns} \label{sec:discussion_toosmall_concerns}

The greatest potential area of concern is that we may not be making an apples-to-apples comparison of galaxy sizes.  In general, one measures $R_{\rm d}$ for extragalactic objects by fitting an exponential model to the disk component of projected surface brightness profiles from imaging.  A variety of ingredients may go into these measurements, such as sky subtraction and inclination correction (see, e.g., H12). Given that it is impossible to measure the integrated light profile of the MW as would be observed externally, photometric estimates of our Galaxy's $R_{\rm d}$ have primarily come from analyzing star counts along various lines-of-sight through the disk (see LN16 for a review of recent measurements).  This requires optimizing a model of the stellar density profile to match observations, requiring assumptions about both stars' initial mass or luminosity function and the effects of dust extinction. While we have ensured that the same basic assumptions about stellar disk structure have been used in both the Galactic and extragalactic estimates of $R_{\rm d}$ utilized in this study, there remain distinct differences in both the nature of the observational data and the analysis techniques used for each. These serve as potential sources of systematic error in our comparisons.

However, as described in LN16 there are a number of reasons to have confidence in the robustness of the Galactic $R_{\rm d}$ value used here. First and foremost, a wide array of mass models for the MW have been developed based upon dynamical (velocity) measurements, which yield estimates of $R_{\rm d}$ that are independent of star count data. These dynamical estimates all fall in the range of $\sim$2--3 kpc, matching well with our estimates of $2.71^{+0.22}_{-0.20}$ kpc and $2.51^{+0.15}_{-0.14}$ kpc based upon visible or IR light, respectively. Furthermore, the observed difference between the visible and IR scale length of the MW is similar to that observed for extragalactic objects.  Based on these facts, it seems unlikely that the MW scale length is different from that measured for typical spirals of the same $L$ and $V_{\rm rot}$ due to a catastrophic error in the Galactic $R_{\rm d}$ measurements.

We could alternatively consider the possibility that our estimate of the MW scale length is correct and instead the value for the Galactic luminosity/mass or rotational velocity is in error. To address this, we have investigated the magnitude of shifts in both $L$ and $V_{\rm rot}$ that are required to push the MW data point on a trajectory toward the $LVR$ relation in PCA space. We find that a $\sim$6$\sigma$ increase in $^0\!M_i$ in combination with a $\sim$6$\sigma$ decrease in $V_{\rm rot}$ could achieve this for the $i$-band luminosity relation, whereas a $\sim$9$\sigma$ decrease in M$_\star$ in combination with a $\sim$6$\sigma$ decrease in $V_{\rm rot}$ accomplishes this for the stellar mass relation. Note that since the MW falls below the $LVR$ relation almost entirely along the $R_{\rm d}$ axis, these shifts also correspond to moving the MW data point roughly along the TFR (and hence the MW would remain highly consistent with that relation). Therefore, it is possible that our Galaxy is truly typical in size for its mass, but this would imply that the current estimates of $L$ and $V_{\rm rot}$ are both biased extraordinarily high in a conspiring way so as to offset each other in the TFR. Given that these estimates are made independently of each other and stem from differing methodologies, this scenario seems highly unlikely.

\subsubsection{External Support} \label{sec:discussion_toosmall_support}

We have found that our Galaxy's intrinsic brightness and internal dynamical motions appear to be linked together just as the extragalactic TFR would predict, suggesting that the relationship between luminous and dark matter in the MW is similar to that in spiral galaxies in general.  In contrast, the MW appears to be unusually compact (and therefore dense) compared to its peers. This may indicate that the MW has followed a distinct and perhaps rare evolutionary path.  Interestingly, a variety of recent studies have come to strikingly similar conclusions by studying the properties of the MW's coterie of satellite dwarf galaxies, which also should be strongly linked to its dark matter halo and formation history.

The most well-known example of this has been coined ``the missing satellites problem'', which signifies the stark contrast between the number of dwarf galaxies (DGs) observationally found in orbit about the MW and the number of dark matter subhaloes (which are expected to host such DGs) that simulations of the Universe with a $\Lambda$CDM cosmology predict for MW-mass galaxies.  This problem has persisted since the early 1990s and has been confirmed by many authors \citep[see, e.g.,][]{Kauffmann93,Klypin99,Moore99,Kravtsov10}, and despite a recent influx of ultra-faint DGs discoveries from studying survey data \citep[e.g.,][]{Willman10,Laevens15,Drlica15}, the fact remains that DGs with $L > 10^6$ $L_\odot$ are $\sim$10 times rarer than expected \citep{Bullock10}.

It also appears that DGs are more compactly distributed around the MW than expected. For example, \citet{Yniguez14} compared spherically averaged radial number counts for $>10^5$ $L_\odot$ DGs within 400 kpc of the MW and its comparably massive neighbor, Andromeda (or M31). They found that MW satellites are much more centrally concentrated than M31's, while also finding that the radial distribution of M31 satellites matches well with predictions from $\Lambda$CDM cosmological simulations of Local Group-like (i.e., MW+M31) pairs.

Finally, the MW also seems to be unusual in that it has two $\sim$10$^9$ $L_\odot$ DGs located within $\sim$60 kpc of it, namely the Small and Large Magellanic Clouds. \citet{Liu11} found that only 3.5\% of MW-mass galaxies in the local Universe fulfill this criterion \citep[see also][]{Busha11stats}. Additionally, \citet{Busha11MCs} investigated the properties of both ``mass analogs'' and ``satellite analogs'' of the MW found in the Bolshoi cosmological simulation data.  This analysis revealed that having not only a MW-like mass but also two Magellanic Cloud-like subhalos associated with them corresponded to host halos having a $\sim$1$\sigma$ higher concentration parameter (i.e., the ratio of the virial radius to the scale radius) and a 60\% larger density of DM within 8 kpc ($\approx R_0$).  Unfortunately, the Bolshoi simulations do not incorporate any luminous matter, so we cannot directly connect this to the distribution of stars (though cf. the correlation between DM halo spin parameter and stellar disk scale length found by \citealp{Burkert15}).  State-of-the-art simulations are now becoming available, however, that model both stars and gas within their parent dark matter halos, and with improving mass resolution \cite[e.g.,][]{ViaLactea,Illustris,Eagle,Wetzel16}. With more and more realistic simulations of MW-like galaxies it should be feasible to investigate whether the anomalies in the MW's scale length and satellite population might be linked, which in turn may improve our understanding of the formation history of our Galaxy.

We also have tested this potential correlation by investigating how M31 fits with the same spiral galaxy scaling relations defined above.  To do so, we adopted the following estimates for M31's properties accompanied by their corresponding studies (parallel to the MW properties presented in Table \ref{table:mw}): total stellar mass, $\text{M}_\star = 10.4\pm0.5\times10^{10}\;{\rm M}_\odot$ \citep[e.g.,][]{Widrow03,Geehan06,Mutch,Barmby07,Tamm12,Sick15}; neutral hydrogen gas mass, ${\rm M}_{\rm HI} = 5\pm1\times10^9\;{\rm M}_\odot$ \citep{Corbelli10}; baryonic mass, $\text{M}_\text{bar} = 11.1\pm0.5\times10^{10}\;{\rm M}_\odot$; rotational velocity, $V_{\rm rot} = 250\pm10$ km/s \citep[see, e.g., Table 3 of][]{Corbelli10}; and photometric disk scale length, $R_{\rm d} = 5.0\pm0.5$ kpc \citep{Geehan06,Corbelli10,Courteau11}.

Promisingly, we find that M31's properties are in excellent agreement with the stellar and baryonic versions of the TFRs and $LVR$ relations investigated here, generically being consistent with them at far below the 1$\sigma$ threshold.  Furthermore, the multiple linear regression model from external galaxies described in \S\ref{sec:methods_lvr} predicts the $i$-band photometric disk scale length of M31 to be $5.04^{+1.95}_{-1.62}$ kpc based upon its $\mathrm{M}_\star$ and $V_{\rm rot}$, and $5.15^{+2.64}_{-1.79}$ kpc based upon its $\text{M}_\text{bar}$ and $V_{\rm rot}$; these are in excellent agreement with the observed value adopted above.

This broadly matches potential explanations for the possible peculiarities of the MW's satellite population. To test this hypothesis more properly, a follow-up study is being carried out using simulation data from \citet{Mao15sim} to explore the correlations between the properties of dark matter halos and the satellite populations that they host (Fielder et al., in preparation).  Empirical tests will also be possible using $LVR$ measurements of MW analogs in the MaNGA sample \citep{Bundy15}, combined with assessments of satellite populations using imaging from the DECaLS survey (\url{http://legacysurvey.org/}, \citealp{Schlegel15,Blum16}).  As cosmological simulations and theoretical machinery continue to improve, the results from this study may ultimately provide strong constraints on how our Galaxy came to be.

\subsubsection{Commentary on Broader Historical Context and Implications} \label{sec:hist_context}
The main conclusion drawn from this study, specifically that our Galaxy is unusually compact in comparison to its peers, provides a significant update to longstanding concerns of whether the MW is typical.  These concerns stem back to pioneering studies of the geometric structure of our Galaxy, and have relevance to one of astronomy's more famous historical debates: the value of the Hubble constant ($H_0$) in the pre-{\it Hubble Space Telescope} era.

The works of \citet[][hereafter dV\&P]{dVP78} and \citet[][hereafter B\&S]{BS1} provided the first two-component (i.e., an exponential disk plus bulge) photometric models of the MW, yielding some of the first realistic estimates of the MW's $R_{\rm d}$, based upon then-recent improvements to star count data and our understanding of the gross morphology of our Galaxy from extragalactic objects.  Initially, the dV\&P model was fit to star counts in the polar caps at $R=R_0$ (with an assumption of $R_0=8$ kpc), indicating a best-fit value of $R_{\rm d} = 3.5$ kpc upon assuming the central surface brightness of the MW to be the \citet{Freeman70} canonical value measured for normal spiral galaxies.  Subsequently, the B\&S model adopted the same value with the authors pointing out that $3.5\pm0.5$ kpc represents the mean scale length of Sbc galaxies in the \citet{Freeman70} sample when $H_0=100$ km s$^{-1}$ Mpc$^{-1}$ is assumed, but that this value becomes three standard deviations from the mean when assuming $H_0=50$ km s$^{-1}$ Mpc$^{-1}$.

To address then-current controversies over the value of $H_0$, \citet{dV83a} introduced the method of ``sosie'' (i.e., the French for doppleganger or look-alike) galaxies \citep{Bottinelli85,dV86}, using the MW as a fundamental calibrator of the extragalactic distance scale in an effort to bypass the perils of other indirect, multistep techniques, the complexities of which could lead to systematic errors in measurements and correspondingly the strong bimodality in $H_0$ estimates.  The premise of the ``sosie''-galaxy technique was simply that when two galaxies are a close match with respect to a multiplicity of calibrated properties (e.g., morphological type, color, bulge-to-total ratio, rotational velocity, etc.), it can be assumed that they are a close match in luminosity and size too --- considered a special case of Hubble's ``principle of the uniformity of nature'' \citep{dV83a}.

In one application of this method, \citet{dV83b} adopted an effective diameter for the MW corresponding to $R_{\rm d} = 3.5$ kpc from the dV\&P two-component model in order to set a benchmark for the local absolute distance scale.  This was directly compared to the effective diameters measured from surface photometry for four of our Galaxy's ``sosies,'' identified using an array of calibrating criteria, finding strong evidence for the ``short'' extragalactic distance scale with a most probably value of $H_0\approx95$ km s$^{-1}$ Mpc$^{-1}$ (as opposed to the rival ``long'' extragalactic distance scale parameterized by $H_0\approx50$ km s$^{-1}$ Mpc$^{-1}$; see, e.g., \citealp{Sandage93a,Sandage93b}).  It is worth pointing out that \citet{dV83a} cautioned that this method relies fundamentally on assuming that our Galaxy is average in all respects; based upon the results presented in this paper, that assumption is not entirely accurate.  We also point out that, to some extent, this reasoning seems circular because the dV\&P model adopted a Freeman-law surface brightness for the Galaxy that was normalized to the short distance scale in order to yield the value of $R_{\rm d} = 3.5$ kpc.

The work of \citet{vdK86} provided an important update to the earlier dV\&P and B\&S models \emph{not} by fitting to new star count data, but by taking advantage of (nearly) all-sky optical imaging of the integrated starlight of the Galaxy from Pioneer 10 --- the first man-made spacecraft to travel beyond the asteroid belt, where the contribution from zodiacal light becomes negligible.  After demonstrating that the Pioneer 10 data could only reasonably constrain the ratio of the exponential disk (radial) scale length to (vertical) scale height, the author adapted the B\&S model to fit the observed surface brightness distribution of the MW after accounting for several known potential sources of systematic error.  The best fitting model provided an estimate for this ratio of $17\pm3$, and multiplying this by the disk scale height estimate of $325\pm25$ pc from the B\&S model (cf. 400 pc for the \citealp{BovyRix13} model, derived from star count decompositions of SDSS \textit{SEGUE} data by \citealp{Bovy2012}) yielded a new estimate for $R_{\rm d}$ of $5.5\pm1.0$ kpc (later revised to $5.0\pm0.5$ kpc; \citealp{vdK87}); in an apparent coincidence, this is in excellent agreement with the predicted MW value from the $LVR$ relation found here.

We note that the \citet{vdK86} disk scale length estimate was most sensitive to the surface brightness profile around $b=20^\circ$ (see Figure 5a of that study) and dependent on a simple dust model \citep{Sandage72} that was somewhat in tension with the data.  Hence, it is reasonable that the treatment of obscuration could have led to systematic errors that biased the $R_{\rm d}$ result high (in tension with the modern value used here at the 2.7$\sigma$ level).  This could be tested by reanalyzing the Pioneer data using the much-improved dust models that are available today, which may allow data at even lower $b$ to be included.

Next, \citet{vdK86} set out to update the ``sosie''-galaxy analysis of \citet{dV83b}, noting that the MW now appears definitely larger than its ``sosies'' are for the short distance scale.  First, the author notes that an update to \citet{dV83b} would lead to $H_0\approx75$ km s$^{-1}$ Mpc$^{-1}$ (comparable to the modern consensus value used here), which appears to be based on the work of \citet{vdKSearle82}.  That study investigated the structural properties of seven spiral galaxies from surface photometry and then directly compared to those of our Galaxy, as well as explored several other lines of reasoning, in order to determine its disk scale length.  This study concluded that the MW is ordinary amongst external spirals only if both the Galactic $R_{\rm d}$ is about 5 kpc and $H_0\approx75$ km s$^{-1}$ Mpc$^{-1}$.  They note that if either the short or long extragalactic distance scale was used instead that the uniformity in spiral galaxy properties shown in their Table 3 would be destroyed.  These values are in excellent agreement with the results we have presented here, and again coincidentally matches well with the Pioneer 10 results.

Extending this method further, \citet{vdK86} directly compared the disk scale length of the Milky Way based on Pioneer 10 data to those measured for external spirals in order to constrain the extragalactic distance scale.  Seemingly taking the MW to be of average size amongst Sb and Sc spirals, the author found that comparisons to three extragalactic samples lead to similar results, which combined for a weighted mean of $H_0=55\pm25$ km s$^{-1}$ Mpc$^{-1}$, whereas comparing independently to a fourth sample yielded $H_0=85\pm25$ km s$^{-1}$ Mpc$^{-1}$.  The author also explored calculating the 26 $V$-mag arcsec$^{-2}$ face-on isophotal diameter of the MW from the Pioneer 10 model (as well as that of M31) and compared to those measured for spiral galaxies in the Virgo cluster.  This lead to estimates of $H_0$ in the range of 60--65 km s$^{-1}$ Mpc$^{-1}$, as well as the realization that if $H_0$ truly is as large as 100 km s$^{-1}$ Mpc$^{-1}$ then the MW and M31 must be extraordinarily large --- in fact, both would be larger than any other spiral galaxy to within a radius of several tens of Mpc.  Ultimately, the debate over $H_0$ would only be settled with the advent of the {\it Hubble Space Telescope}.

Taking a different approach, the study of \citet[][see also \citealp{Karachentsev96}]{Goodwin98} set out to test the hypothesis that the MW is an \emph{overly} large spiral galaxy as posited by earlier works (e.g., \citealp{Shapley72}; dV\&P; \citealp{Rubin83}; \citealp{vdK90_sb}).  Noting that the aforementioned short extragalactic distance scale implies that the MW must be an unusually large system, whereas the long distance scale implies it to be very ordinary, \citeauthor{Goodwin98} set out to bypass any systematic errors tied to $H_0$ by using Cepheid-based distances to 17 nearby galaxies to place the MW in perspective.  Here, the authors converted the \citet{vdK86} Pioneer 10 results into an estimate for the MW's 25 $B$-mag arcsec$^{-2}$ isophotal diameter ($D_{25}$) and compared directly to those of the external systems, finding the MW to lie almost exactly on the mean of galaxy sizes (see their Figure 1).  Next, they investigated the effect of comparing $D_{25}$ as a function of morphological type $T$ as recorded in the \textit{Third Reference Catalogue of Bright Galaxies} \citep[RC3;][]{RC3}.  Taking the best estimate of the MW's morphological type to be $T=4$ (corresponding to Sbc, in excellent agreement with \citealp{dV83a,Licquia1}), and bracketing this value between two and six (i.e., Sab and Scd), the authors restricted their extragalactic sample to just 12 galaxies that lie in this range.  They found that the MW's $D_{25}$ fell somewhat below but still well within one standard deviation of the sample mean (see their Figure 2).  Overall, this lead them to conclude that our Galaxy is \emph{not} one of the largest galaxies, but is very averagely sized for its Hubble type.

We can update this comparison here using our estimate for the Galactic $R_{\rm d}$ and more extensive extragalactic data by incorporating $T$ values from RC3, which may be obtained for each galaxy from the SFI++ catalog.  To do so, we have restricted our fiducial sample of 258 moderately-inclined spiral galaxies to just those with $2\leq T\leq6$, yielding a sample of 240 objects, and then measured the mean and standard deviation of their $R_{\rm d}$ values, finding these values to be 3.87 and 1.97 kpc, respectively. Our current estimate for the Galactic $R_{\rm d}$ of $2.71^{+0.22}_{-0.20}$ kpc falls below the mean value (and is inconsistent with it at the $\sim$4.5$\sigma$ confidence level), but still lies well within one standard deviation of it.  If we are to restrict more narrowly to objects with $3\leq T\leq5$, corresponding to a sample size of 230, then we find that repeating this analysis leads to very similar results.  Therefore, we can confidently deem the MW to be below-average in size for its Hubble type, though not as drastically as for its $L$ and $V_{\rm rot}$.

Finally, based on our finding that the MW is extraordinarily compact given its mass/luminosity and rotational velocity, which rests upon modern, independently determined estimates of the MW $R_{\rm d}$ and $H_0$, we infer that the MW must be an extraordinarily high surface brightness system and should have an unusually maximal disk \citep[cf.][]{Sackett97,CourteauRix99,Bershady11,BovyRix13}.  The former hypothesis can be tested further using the methods of \citet{Licquia2}, which may be considered an update to the method of studying ``sosie'' galaxies.  This would entail identifying a sample of spiral galaxies whose distribution of both absolute magnitudes (or alternatively stellar masses and star formation rates, as was used in that original study) and disk scale lengths match the posterior PDFs describing those properties for the MW.  The corresponding distributions of central surface brightnesses and isophotal diameters for such a sample of Milky Way analogs could then be used to constrain the values that are plausible for our Galaxy.

\section{Summary and Conclusion} \label{sec:summary}
Overall, our results provide a significant improvement in our understanding of how the MW fits in extragalactic contexts.  First and foremost, with consistently normalized Galactic and extragalactic data, we find excellent agreement of MW properties with TFRs measured for other spiral galaxies, in contrast to prior investigations, which deemed our Galaxy a 1--1.5$\sigma$ outlier to the relation. These tensions can be almost entirely explained by systematic errors in past work, particularly the values assumed for the distance of the Sun from the Galactic center ($R_0$) and for the $R_{\rm d}$ of the MW, which both affect luminosity estimates. The Galactic luminosity estimates used here are based upon updated knowledge of these structural parameters and incorporate all associated uncertainties. This overall conclusion holds up when comparing our MW estimates to TFRs found by other authors, where systematic differences between Galactic and extragalactic measurements could be a greater problem; nonetheless, the MW falls sometimes above or below these relations, and is generically consistent with them at below with 1$\sigma$ threshold. We can confidently deem the MW a typical galaxy in the context of the TFR, and hence it is a suitable laboratory for studying the driving mechanisms of the relation.

By extending this type of comparison to 3-dimensional $LVR$ relations, we also have established strong evidence that our Galaxy appears unusually compact for its mass. The MW disk would need to be nearly twice as large as the measured value to be of typical size given our Galaxy's $L$ and $V_{\rm rot}$.  Furthermore, we find that $\sim$90\% of spiral galaxies lie closer to the $LVR$ relation in 3D parameter space than does our Galaxy, characterizing it as a rather substantial outlier in this context, though not extraordinarily so.  We infer from our findings that our Galaxy also must be a high surface brightness system and should have an unusually maximal disk given its concentration.  Many simulations of galaxy formation have been tuned to yield MW-like scale lengths on average for galaxies that resemble the MW in other respects \citep[e.g.,][]{Governato08,Bonoli15}; however, the implicit assumption made that the MW is typical in its disk properties clearly fails.  This broadly matches suggestions that the MW may be ``too small'' based upon its satellite population. New studies of state-of-the-art simulation data are underway to investigate the potential correlation between a galaxy's size and the properties of its satellite population.  Ultimately, the results presented here may serve as a key step in understanding the distinct evolutionary history of our Galaxy.

\section*{Acknowledgments}
We are grateful to Chad Schafer for sharing with us his expertise on some of the statistical analysis techniques we employed. We thank Andrew Zentner for several useful discussions about the ideas presented herein. We also thank the anonymous referee for providing helpful suggestions. TCL and JAN are supported by the National Science Foundation (NSF) through grant NSF AST 08-06732, as is MAB through grant NSF AST 15-17006. TCL also acknowledges support from a PITT PACC fellowship, and an Andrew Mellon Predoctoral Fellowship. This research made use of the IPython package \citep{ipython} and SciPy \citep{scipy}. Funding for SDSS-III has been provided by the Alfred P. Sloan Foundation, the Participating Institutions, the National Science Foundation, and the U.S. Department of Energy Office of Science. The SDSS-III web site is http://www.sdss3.org/. SDSS-III is managed by the Astrophysical Research Consortium for the Participating Institutions of the SDSS-III Collaboration including the University of Arizona, the Brazilian Participation Group, Brookhaven National Laboratory, University of Cambridge, Carnegie Mellon University, University of Florida, the French Participation Group, the German Participation Group, Harvard University, the Instituto de Astrofisica de Canarias, the Michigan State/Notre Dame/JINA Participation Group, Johns Hopkins University, Lawrence Berkeley National Laboratory, Max Planck Institute for Astrophysics, Max Planck Institute for Extraterrestrial Physics, New Mexico State University, New York University, Ohio State University, Pennsylvania State University, University of Portsmouth, Princeton University, the Spanish Participation Group, University of Tokyo, University of Utah, Vanderbilt University, University of Virginia, University of Washington, and Yale University.


\end{document}